# A mobile digital device proficiency performance test for cognitive clinical research

(Short title: Mobile digital device abilities test for cognitive clinical research)

*October 5th, 2024 (updated version)*

Alan **Cronemberger Andrade** [1]* ; Diógenes de Souza **Bido**[2] ; Ana Carolina Bottura **de Barros**[3] ; Walter Richard **Boot**[4] ; Paulo Henrique Ferreira **Bertolucci** [1]

[1] Department of Neurology and Neurosurgery, Universidade Federal de São Paulo, São Paulo, Brazil

[2] Center for Social and Applied Sciences, Mackenzie Presbyterian University, São Paulo, Brazil

[3] Centre for Neuroscience, University of Glasgow, Scotland, United Kingdom

[4] Attention and Training Lab, Department of Psychology, Florida State University, Tallahassee, United States

**\* Corresponding author:**

Alan Cronemberger Andrade

Department of Neurology and Neurosurgery, Universidade Federal de São Paulo

Rua Pedro de Toledo, 650 - CEP 04039-002 - Vila Clementino, São Paulo

Phone: +55 (11) 5089-9200

Email: alan.andrade@unifesp.br

# Abstract

Mobile device proficiency is increasingly important for everyday living, including to deliver healthcare services. Human-device interactions represent a potential in cognitive neurology and aging research. Although traditional pen-and-paper evaluations serve as valuable tools within public health strategies for population-scale cognitive assessments, digital devices could amplify cognitive assessment. However, even person-centered studies often fail to incorporate measures of mobile device proficiency and research with digital mobile technology frequently neglects these evaluations. Besides that, cognitive screening, a fundamental part of brain health evaluation and a widely accepted strategy to identify high-risk individuals vulnerable to cognitive impairment and dementia, has research using digital devices for older adults in need for standardization. To address this shortfall, the DigiTAU collaborative and interdisciplinary project is creating refined methodological parameters for the investigation of digital biomarkers. With careful consideration of cognitive design elements, here we describe the open-source and performance-based Mobile Device Abilities Test (MDAT), a simple, low-cost, and reproductible open-sourced test framework. This result was achieved with a cross-sectional study population sample of 101 low and middle-income subjects aged 20 to 79 years old. Partial least squares structural equation modeling (PLS-SEM) was used to assess the measurement of the construct. It was possible to achieve a reliable method with internal consistency, good content validity related to digital competences, and that does not have much interference with auto-perceived global functional disability, health self-perception, and motor dexterity. Limitations for this method are discussed and paths to improve and establish better standards are highlighted.

# Introduction

Mobile digital devices proficiency is becoming increasingly important. The contemporary lifestyle, characterized by the omnipresence of digital mobile devices, is shaping daily living, altering the way people think, interact, and process information. In 2021, 97% of adults in developed countries already used smartphones. Within emerging economies, in 2019, smartphone ownership rates ranged from highs of 60% in South Africa and Brazil to around 40% in Indonesia, Kenya, and Nigeria (Pew, 2019, 2021) So, due to inequalities, not all users may be benefiting equally from digital advancements. Many older adults also do not own or use these devices, resulting in a proficiency gap. In Brazil, between 2019 and 2021, internet use rose from 34 per cent to 48 per cent, respectively, among people aged 60 and over, but this is by far the portion of the population that uses the internet the least (CGI.br, 2023). These digital divide gaps need to be understood and considered, especially for healthcare delivery using these types of devices, including in the assessment of cognitive status.

While mobile device proficiency has become a social requisite, the extent of implications on cognitive behavioral and health are to be determined. Clearer beneficial uses for healthcare are adherence to treatment, screening, asynchronous monitoring, and remote access of mental health care resources for underserved communities (Chan et al., 2022; Ravindran et al., 2023; Sugarman & Busch, 2023). The devices have undeniably enriched our abilities, but their ubiquity raises health concerns: altering sleep patterns, bringing cognitive load and stress (technophobia), reducing face-to-face social interactions, and also being related to mood disorders (Ophir et al., 2009; Primack et al., 2021; Sheppard & Wolffsohn, 2018). Considering older adults and evidence, interpretation of results on how older adults interacts with these technologies show that there are mainly technical, physical/cognitive, and socioeconomic barriers (CGI.br, 2023; Charness & Boot, 2022; Nishijima et al., 2017).

Despite some hurdles, there is research going on with good results. Digital mobile cognitive assessments already has concurrent validity with neuropsychological tests for quite some time (Allard et al., 2014). Computer-based cognitive assessment is a reality in clinical trials (Papp et al., 2021); virtual reality strategies are being tested (Zygouris et al., 2017), and gamified strategies are being progressively improved (Vermeir et al., 2020). There is even more research about how one could monitor cognitive function, an important component for dementia staging, by self-reported online instruments (Nosheny et al., 2020). Although, caution is warranted. Improved methodological validation steps are needed.

One of these steps, digital device proficiency – sometimes referred as "digital literacy" – is not assessed or reported in many digital mobile cognitive screening studies (Mansbach et al., 2020; Öhman et al., 2022). Of 188 studies analyzed, researchers found that only 29.2% of health technology research for older adults incorporate any kind of technology proficiency/familiarity measures, only 5.8% report computer/internet usage, and only 1.6% report technology ownership (Harrington et al., 2022). To decipher connections between mobile device proficiency and cognitive abilities, methodological rigor is essential. Researchers must differentiate between simply being prone at using a device from deeper levels on how that person's proficiency integrate with daily skills, beyond age differences (Carr et al., 2022). It's vital to factor in variabilities like user age, education, social context, and the nature of device interactions, ensuring conclusions drawn are accurate and meaningful (Czaja et al., 2006). Different sociodemographic profiles could also determine different costs and access to a correctly personalized and cost-effective approach. Barriers to the adoption of online mass testing of people with cognitive problems can occur due to difficulties in accessibility and digital proficiency.

However, few *objective* measures of mobile device proficiency exist. Finding an ideal proficiency assessment to measure complex constructs such as functional technology usage for a single or grouped activities is difficult. There is no gold standard test for measuring digital mobile device proficiency, and each one has individual characteristics (Oh et al., 2021). Questionnaire-based psychometric instruments such as the MDPQ and eHEALS can be used, but are self-report and not performance based (Norman & Skinner, 2006; Roque & Boot, 2018). Despite this, as questionnaires have good validity and practical use techniques then can provide a level of utility (Petrovčič et al., 2019). Even some performance-based measures are not sufficient. Another way of assessing this proficiency, for example, is looking to multiple competences at given timeframe, which have broader usage in the corporate sphere and in education, and even for technical workers, but are not suitable for simple clinical usage. Indirectly, it is even possible to look to mobile device app usage and try to infer a person level of proficient usage or cognitive ability (Gordon et al., 2019).

This is all especially important for detecting cognitive problems in older adults, where subtle changes may be prone to be detected with digital means. Mild cognitive impairment (MCI) has been considered a clinical condition that can precede a dementia syndrome, and there is measured potential for its timely diagnosis to reduce the risk of dementia (Rovner et al., 2018). In large clinical studies such as the ADNI cohorts, around 33 per cent of patients with the condition will develop dementia within 3 years (Basaia et al., 2019; Chen et al., 2022). Its diagnosis is already a reality in specialized centers, with some reporting more than two decades of experience (Glynn et al., 2021). However, the problem is not always easy to diagnose. There are barriers in contexts without adequate technological resources, trained personnel, multi-professional support, high costs, and complex uncertainties (Akinyemi et al., 2022; Michalowsky et al., 2018; Swallow, 2020). Before using new possibilities for diagnostic screening of MCI - and recent advances in biomarkers in cerebrospinal fluid, serum plasma, and neuroimaging - selection of high-risk subjects seems to be an important complementary strategy (Hansson et al., 2018; Karikari et al., 2021; La Joie et al., 2019). At this earlier point in the clinical course of Alzheimer's disease – a milder stage before dementia sets in – new non-pharmacological and pharmacological therapeutic interventions are most often proposed (Sims et al., 2023; van Dyck et al., 2023; Whitty et al., 2020). This is where there is potential to find high-risk individuals using digital mobile technologies, as mass cognitive testing to detect early signs, for example.

Having low-cost but high-tech strategies is a necessity for screening. The period around the diagnosis of Alzheimer's disease is very costly, with 91% increase in *total disease costs* around "year zero" (Sopina et al., 2019). A great demand in the field is to offer greater availability of neuropsychological tests to patients to be trialed and followed up, who can have their neurological problems affecting cognition preferably identified in primary care setting. The best tests have good sensibility (around 83 to 97%) and specificity (70 to 100%) on systematic studies, but these features are usually not present on the same tool (Zhuang et al., 2021). Efforts to adopt cognitive testing *en masse* and early should be accompanied by test standardization and harmonization, which is already taking place for conventional pen and paper tests. Favorable clinicians report finances (15%) and digitalization (9%) as facilitating measures to implement uniform testing in Europe (Boccardi et al., 2022; Grazia et al., 2023).

To fill these gaps, our research group has, in collaboration with experts in technology and aging, have developed a short, easily administered measure of mobile device proficiency that can benefit the care of older adults, and by accurately understanding mobile device proficiency, directing benefits to older adults. This is crucial for understanding how to deliver interventions, additional training and support some older adults might need, and how to understand data collected from these devices. The complex interplays around human-device interactions, cognitive neurology, usability, personal abilities, and cultural relations with technologies do not have a clear answer. The aim of setting the DigiTAU research is to improve methodological standards, especially validating psychometric measures for digital health and cognitive impairment research. The project uses refined methodological parameters for the investigation of digital biomarkers for cognitive research using digital devices. By exploring connections with psychometric parameters, the current study offers a testing framework for mobile device proficiency, the Mobile Device Abilities Test (MDAT), an open-source performance-based approach. The hypothesis was that this competence-based measurement framework would be valid in terms of its construct and content and would differentiate proficiency generational gaps.

# Methods

## Study participants

The sample was composed of 101 healthy adults, with no detected cognitive impairment, from the DigiTAU cross-sectional study population, who completed the MDAT test. They were recruited in the community (São Paulo metropolitan region) throughout January 2021 and May 2023, using wall posters, institutional websites, and non-paid social media publishing. The DigiTAU project is an initiative based at the Behavioural Neurology Section of the Department of Neurology and Neurosurgery Federal University of São Paulo (Unifesp) to study digital biomarkers in Brazilian populations. It involves methodological pilot studies of digital biomarkers of cognitive impairment and dementia for clinical, research, and telehealth care (CRONEMBERGER ANDRADE *et al.*, 2020, 2021). The study protocol was registered *a priori* on the Brazilian government-sponsored human health studies structured online repository *Brasil Platform* (plataformabrasil.saude.gov.br). It was approved by Unifesp Institutional Review Board (CEP-Unifesp Project No. 0722/2020), in accordance with international law, and followed established standards as stated in STROBE Statement checklist for observational studies (von Elm et al., 2007). Formal signed consent was obtained prior to enrollment for all participants.

**Inclusion and exclusion criteria.** We included individuals from 20 to 79 years old that were considered cognitively functional, split between age groups, with the following inclusion criteria: (a) no history of major neurological or psychiatric disease, (b) no current use of psychotropic medication, (c) individuals that came alone and by their own interest and means, (d) said no to cognitive health problems or cognitive functional disabilities, not accounting for self-reported memory complaints. People with major motor, auditory, visual deficits, as well as those with no previous contact with mobile digital devices, were excluded during triage.

All the tests were done by a trained neurologist. Initially, a total of 341 people were individually contacted and trialed, 138 booked appointments, 105 attended the sessions, with a no-show rate of 23,9%. Out of these, 4 participants had chosen to discontinue their participation after attending (two left earlier due to job or study motifs, one felt uncomfortable being tested, and one received an emergency call). A total of 101 participants completed the evaluations. All subjects resided in São Paulo, Brazil, though many originated from other regions (31.7\%). The subjects were assessed for the absence of dementia and then had self-reported health information, daily functionality, and cognitive parameters measured. See Results section and Table 1 for sample characteristics.

**Mobile device test description**

To test abilities of mobile digital device usage it was developed a performance-based semi-ecological (i.e., an ecological measure of a simulated setting) test. The intention was that the Mobile Device Abilities Test (MDAT) framework could provide a short performance test setting (less than 2 to 5 minutes to finish), easy to perform, with clear instructions, understandable to the participants, including a mix of very basic, but essential daily functional activities for a cognitively healthy person. The construct to be measured with this test was how proficient a cognitively healthy person is in relation to mobile digital devices, focused on brain health. The aim was to have this short, ecologically valid, and reliable test to compare it with other measures, specially focusing on mobile health interventions in cognitively healthy people or people with mild pre-dementia symptoms for future studies. The context of use is directed to checking proficient mobile device usage for health interventions, using digital devices as a methodological standard. The target population is represented in the sample, following methodological recommendations to avoid bias (McKenna & Heaney, 2021; Mokkink et al., 2010).

First, after examining a set of instructions or possible items reviewing the literature, it was noticed that the whole set of different activities or competences using a mobile phone in our daily lives is very large. There was a need to select a small pool of relevant items to be included at the MDAT. An UNESCO report examining nine country-level digital proficiency/literacy competence frameworks – mainly directed to education or business-related digital proficiency – cite on average 79.2 mapped instances (ranging from 12 to 177 different ones), grouped in 5 to 7 competence areas (Law et al., 2018). This amount is clinically inviable to test. Previous mental health research included as many as 46 different activities grouped in 8 competence areas using mobile devices in a questionnaire (Roque & Boot, 2018). The final activities choice the MDAT would cover was based on empirical assumptions that a good ecological test should have the commonly cited basic competence areas explored as task-related MDAT items: the tasks should be relevant as shown in the literature. The selected ones would represent at least how a participant can perform activities related to: (a) privacy/safety skills (unlock with a password), (b) communication (send a message), (c) entertainment/content creation (take a photo), and (d) basic operational activities (turn off the phone). These tasks final selection was mainly based on relevance and the similarity between the domain areas also included at digital competence frameworks (Carretero et al., 2017). Deeper examination of the selected tasks is detailed in "content analysis" section.

For this study, a Galaxy® S7 (Samsung Group, South Korea) mobile phone device, a consumer-grade SM-G960F international model with Android® (Google LLC, USA) 8.0 operating system (OS), was utilized for conducting the human computer interaction (HCI) experiments. Announced as a higher end model in Mar 2016 (Gibbs, 2016), the device was chosen for its technical features, and for maintaining technical features compatible with some 2021-2022 mid-price range launched mobile phones sold in the country. It had a 5.1-inch AMOLED 2560 x 1440 pixels resolution display, an ARM® Cortex-A53 octa-core processor, Mali-T880 MP12 GPU, 4GB of RAM, gesture recognition, multi-touch inputs, and pressure sensitivity. Sensors such as an accelerometer, gyroscope, and barometer allowed enough tracking of user interactions. These specifications were deemed necessary for the HCI tasks performed. All participants utilized the same device, and no changes in its operating system or applications were made during the experiments.

In the experimental design, the sequence of tasks was intentionally fixed, guided by some considerations: (a) the brief duration of the test minimized the potential for fatigue effects on participant performance; (b) preliminary observations indicated no significant fatigue, supporting this methodological choice; (c) from a neuropsychological perspective, the impact of task order would be minimal within the MDAT; (d) standardized the task sequence ensured consistent test conditions across all participants, essential for eliminating variability in test administration as a confounding factor; and (e) a fixed task order could facilitate a cognitively smooth transition between tasks, enhancing the overall participant experience, quality, and reliability of the collected data. The rationale behind this design choice, aligned with practical considerations and specific objectives of the study, suggested that while task randomization would be a valuable approach, it was not considered the best choice.

The initial screen for the "unlock task" (*Task 1*) was adapted to Google Now Launcher 1.4.large (Google LLC, USA), a Google official screen launcher running on Android 4.1 (Kahn, 2014). The "message task" (*Task 2*) was performed using the installed custom/stock Samsung Messages free application, version 4.4.30.59, from Feb 2018 (Samsung Electronics Co., Ltd., South Korea). The "photography task" (*Task 3*) was performed using the free and open-source software, Open Camera app, for Android™ phones and tablets (Harman, 2013), version 1.49.1 (Sep 2021) and 1.49.2 (Jan 2022). The software has advanced features, such as auto-levelling, exposure control, timer, remote control, GPS location tagging, overlay, supports Camera2 API, burst mode, RAW files, slow-motion video, log profile video, noise reduction, dynamic range optimization modes, on-screen histogram, zebra stripes, and focus peaking. The application is offered at no cost, and the source code is available. The "turn-off task" (*Task 4*), was the last one. The test setting is described in Figure 1.

The test instructions were given once and could be repeated (see Supplementary Material 1 and Supplementary Material 2 for details and full text of current instructions for both Portuguese and English language tests, respectively). The instructions sets are available as online resources. All test here described was done only in its Brazilian Portuguese version and would need proper validation in English.

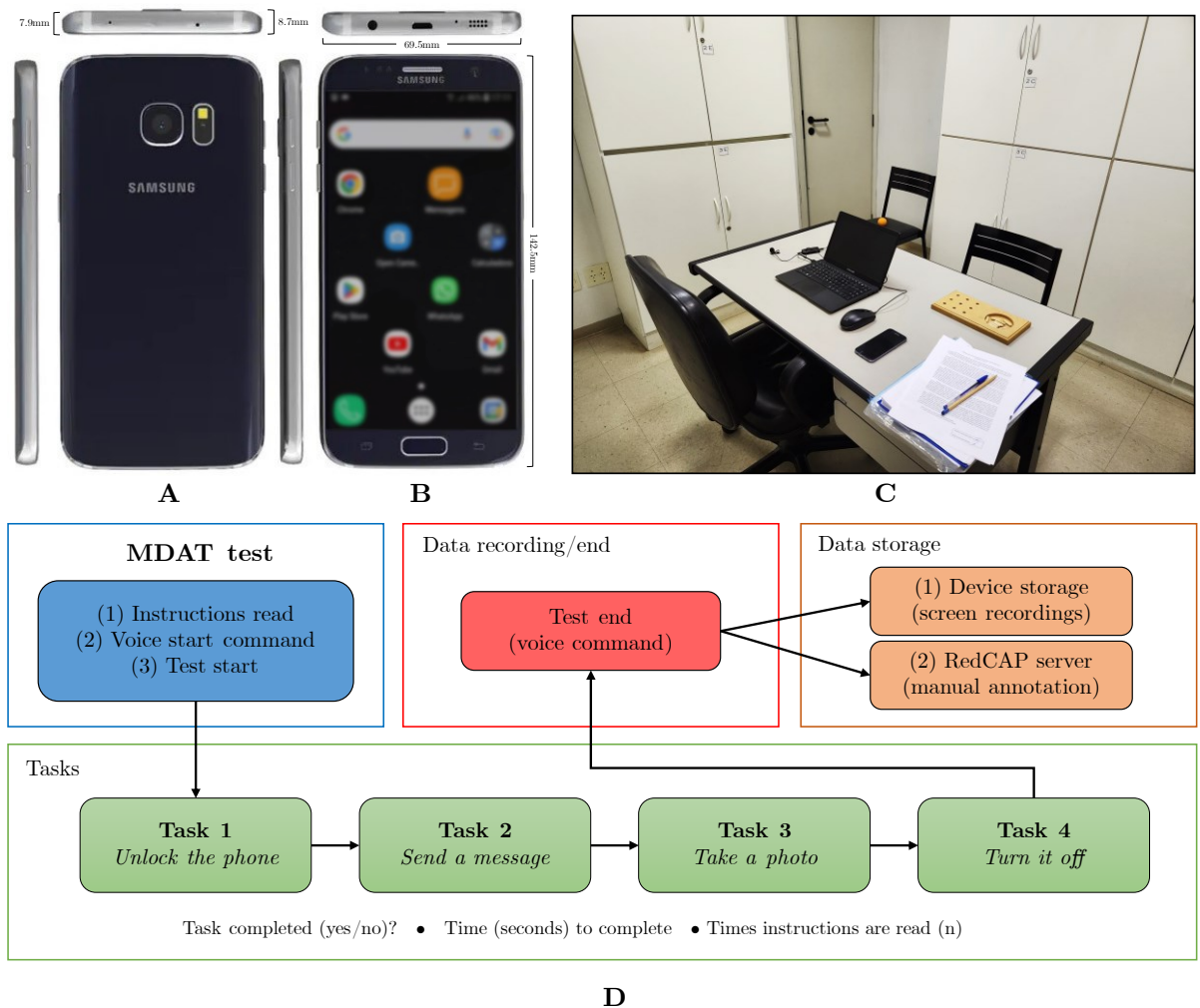

**Figure 1.** MDAT test framework description. At the upper left corner, the smartphone size specifications (A) and initial screen visualized by the participant (B). The e-mail app and the messaging apps other than custom SMS messaging app were deactivated and are "fake" buttons, on purpose. Proprietary apps icons were blurred. Test room (C) shows how was the spatial distribution of the participant testing room configuration. Note the tennis-sized orange ball at a chair behind the participant, and the 9-HPT pegboard. (D) Test execution description. The tasks are executed at the specific order shown here. Voice commands are given at each step and screen were recorded for further examination. Android, Google Play, and the Google Play logo are trademarks of Google LLC. Samsung logo and Samsung Galaxy are registered trademarks of Samsung Electronics Co., Ltd.

## Motor tasks measurements

The participants had their hands motor skills assessed for hand dexterity using the 9-pin hole test (9-PHT), measured for both hands. The 9-PHT is a quantitative assessment tool employed in evaluating fine motor skills and manual dexterity, particularly in the context of clinical and experimental research (Grice et al., 2003; Mathiowetz et al., 1985) and is related to Purdue pegboard test (Tiffin & Asher, 1948). The 9-PHT measures participants' ability to grasp, manipulate, and release small pegs within a specified time frame, which provides insight into their hand-eye coordination and precision. Findings reveal impaired dexterity on 9-HPT at each worse Parkinson's disease (PD) stage, with poorer performance in advanced PD, and significant correlations between pegboard performance and self-reported hand function (Proud et al., 2021). Both were taken as a one-trial data gathering. If major error or misunderstanding implicated the trial (for example, if the participant let the mobile device fall or exit the application accidentally), a 3-trials average was taken. This occurred in less than 5% of the measurements.

**Content and construct validity analysis**

Validity of the MDAT test was assessed in a series of different analyses. At first, some assumptions were made presuming that MDAT models should concurrently correlate with other indicators of mobile usage, especially how much time that individual was using mobile digital devices (A), and its daily usage average time (B). It was also hypothesized that, if it is a valid measure of proficiency, it should be sensible to differentiate generational gaps, so that it would perceive different age groups differently: age *group 1* (from 20 to 39 years), *group 2* (40 to 59 years), *group 3* (60 to 79 years).

**Content validity.** Content validity concerns item sampling adequacy or the extent to which a specific set of items reflects a content domain (DeVellis & Thorpe, 2021). So, we asked to 7 experts' panel (a PhD computer scientist professor, a dementia neurologist, a psychiatrist, a geriatrician, an adult neuropsychologist, a PhD occupational therapist, and a UX/UI professional developer/programmer) to evaluate the MDAT tasks/instructions.

To ensure the quality and ethical appraisal, the expert group members were selected based on specific criteria: (a) each came from different institutions and had varied expertise backgrounds (diversity of perspectives); (b) four of the seven members were women (gender representation); (c) none had direct involvement with the project or any hierarchical or commercial relationships with the authors (independence); (d) all agreed to participate voluntarily and were informed beforehand that no bonuses or consulting fees would be provided (evaluation's integrity); (e) members were anonymous to each other (potential biases from pre-existing relationships reduced); (f) each had previous experience in scientific evaluations (appraisal expertise); (g) their interviews were electronically recorded and transcribed (transparent and accurate documentation); (h) interviews were planned, scheduled in advance, and designed to less than 2 hours duration (focused judgment optimization).

To initially evaluate the content validity for the instrument the groups of experts were asked about the task's text (a) language clarity, inferred (b) text comprehension by the participants, (c) item adequacy practical relevance, and (d) item comprehensiveness relevance for the whole theoretical test set. The group was given a form to evaluate, using a 4-points Likert scale system, the 6 task's instructions "blocks" – the instructions sets for task-1 to task-4, total time instruction, and task repetition instruction. After they received equal verbal explanation and time to clear doubts, they were interviewed and replied to a structured form about their impressions, which were done through an online video call, synchronously, recorded, electronically transcribed. The general content had no major qualitative objection, and 3 of the 6 instruction blocks had more than 40% of the experts suggesting text improvements, which were all accepted. But, to also give a quantitative measurement of the content validity, it was calculated a Content Validity Index (CVI) for both individual items (i-CVI, not shown) and whole scales (S-CVI), following recommendations with a Likert-type scoring system (very poor, poor, good, very good), scoring a judge positive feedback when good/very good answers were given (Polit et al., 2007; Alexandre & Coluci, 2011). An endorsed CVI of 0.86 or higher was considered to indicate a good scoring for each item characteristic and scales based on recommendations for a 7-experts panel (Lynn, 1986).

To have a matched analysis of the selected tasks, they were also subject to appreciation by the panel group of experts about how well they would fit the competence areas included in DigComp 2.2. This framework points essential digital skills that people need to have to manage nowadays needs. It is derived from an European Commission study group, and describes current digital functional usages, that are all included in 5 competency domains: information and data literacy (1), communication and collaboration (2), digital content creation (3), safety (4), and problem solving (5), also subdivided in more specific competence tasks (Vuorikari et al., 2022). Examination of content domains and correlation between digital modern competencies items of the MDAT tasks using the DigComp framework was done in similar to other research (Oh et al., 2021), but here was done by asking the experts how MDAT items fit and related to this areas, using a 0-to-100-point scale, for each of the competence domains covering the constructs (see results in Table 2).

**Construct validity.** Construct validity refers to the degree to which an instrument measures the intended theoretical construct or concept. Different techniques to evaluate the construct were employed, mainly to determine latent variable structures. To assess the construct validity of the proposed psychometric instrument, a non-traditional factor analysis was conducted, following published theory to do structural equation modeling with partial least squares estimation (Hair et al., 2021; Kline, 2023). The following section details its features.

**Measurements of digital device usage**

All participants were asked how often they used computers and/or mobile digital devices (tablets or smartphones) and classified either as daily or non-daily users (use of any smartphone or tablet digitals device at least once a day) and were also categorized for the time in years and current daily usage hours. To check if different operating usage was much different between young and older participants, we asked which mobile model they were mostly using in a daily basis and calculated the statistical difference between operating system for individuals above 50 years old versus younger. Similar comparative methods were used before for research in mobile device proficiency (Roque & Boot, 2018).

**Mathematical modeling for reliability and validity analysis**

At first, before a structural equation modeling (SEM) approach was employed, alternative models were tested using dichotomization of the MDAT task's variables, converting continuous or ordinal into binary variables based on predefined thresholds. It was done mainly to dichotomize total time and instructions repetitions. This offers some advantages to simplify relationships, interpretation, and reduce outliers influence, but may cause information loss, reduced statistical power and add potential bias, so after this study step it was considered not ideal (Cohen, 1983) as it could lead to potential misrepresentation of the relationships among variables due to the elimination of variability within the data (MacCallum et al., 2002). Models results using dichotomized variables were compared with results obtained by using the original continuous variables. Dichotomization was an easy method to calculate and give scoring, but in fact led to a poor result. This comparison was employed to understand models' sensitivity to the scale of measurements for usage time/contact.

There is solid formative theory with studied models that led us to decide on the *a priori* relationships between the variables in the study (Czaja et al., 2006; Lee et al., 2019). Because the variables had mapped relationships with digital proficiency, with known number and meaning of the factor, this study did not use exploratory factor analysis techniques as these assumptions would make the models inappropriate (Jöreskog, 2007). Factor analysis was not preferred in this case, as it was required an approach aimed at: (i) predicting external variables (device usage), (ii) analyzing a smaller sample due to complex specificities, (iii) looking into data did not follow a normal distribution, (iv) the need for a data-driven component weights maximization through a determinate structure rather than factor-driven, (v) implying that measurement errors were unlikely (a common pre-analytical assumption in factor analysis). Robust polychoric and polyserial correlation matrices, suitable for these set of variables, could make a factorial MDAT model error-prone (Rigdon & Ferguson, 1991). Partial least squares structural equation modeling (PLS-SEM) can handle complex high order models with many constructs and structural paths and has complimentary benefits when compared to covariance-based confirmatory factorial analysis (Rigdon et al., 2017; Willaby et al., 2015). To do this, all the variables were also transformed into the same directionality, i.e. higher values implying a direct correlation with greater proficiency in digital devices usage.

Then, in the next step, it was hypothesized that a latent variable (*usage_time*) containing information of the participants time in years (*time_years*) and the current usage per day in hours (*daily_usage*) would be able to explain the proficiency of the individuals, using linear regression statistics. So, in the end , the best model would be the one that has this explanatory capability (1) and internal consistency (2), as main features. To analyze the relationships between latent variables, and to test the proposed mathematical model, we employed PLS-SEM using SmartPLS software Version 4 (Ringle et al., 2015). This method is particularly suited for exploratory research and analyzing complex models with multiple variables (Hair et al., 2021). The continuous variables provided a more accurate representation at the PLS-SEM model. Internal consistency of the MDAT items was calculated using this SEM method also to explore scale dimensionality. Variables within latent constructs were altered to assess their impact on the overall models. Different interpretations of the underlying constructs were done to ensure that the findings were not reliant on single conceptualizations. Table 5 at the results article section shows the different models' characteristics.

**Statistical analysis**

Data was analyzed using JASP (JASP-Team, 2023), R Statistical Software (R-Core-Team, 2023), and SmartPLS (Ringle et al., 2022) software, based on their distinct analytical strengths and use cases. The data were initially subjected to descriptive analysis. Categorical variables were quantified as frequencies and percentages, while continuous variables were either stated as mean ± standard deviation for normally distributed data, or median with interquartile range (IQR) for non-normally distributed data. The normality of the data distribution was verified with Shapiro-Wilk test, and Q-Q plots inspection.

Subsequently, inferential statistical methods were applied. For comparisons of more than two groups, ANOVA or Kruskal-Wallis's test were employed, based on the normality of data. The Chi-squared or Fisher's exact tests were selected as appropriate for categorical variables. To measure the strength and direction of relationships between two continuous variables, Pearson's, Spearman's, or Kendall's correlation coefficients were calculated in accordance with data distribution (Kendall, 1938). Linear and logistic regression models were applied to estimate the influence of multiple independent variables on the dependent variables.

Given that the MDAT assessed mobile digital devices proficiency in different age ranges, the modeled scale items would show a pattern such that individuals from different generations (*age groups*) would have different abilities, with older generations with less proficiency. The evaluation of different self-perceived disability was done with WHO-DAS 2.0 12-item sum score (Ustün et al., 2010). The EQ-5D-5L instrument, an assessment of health-related quality of life, was used to measure health perception during the experiment (Buchholz et al., 2018).

It was then hypothesized that: (1) self-reported functional status measured by the WHO-DAS instrument was not statistically different between groups, (2) there should be differences for MDAT scores between the age groups, suggesting a proficiency generational divide, (c) no group differences in health status measured with EQ-5D-5L would be noticed. The absence or presence of a statistically significant difference in the context of the hypothesis would not imply definite equivalence/difference between groups, but could indicate inferences within the constraints of this study sample. The ANOVA test was applied to compare means among the groups.

The validation for consistency among the scores attributed by the 7 experts' group was achieved through the statistical computation of intraclass correlation coefficients (ICC). It served to validate consistency among their evaluations (a) and allowed subjective interpretation of the content parameters to be assessed in a more robust, quantitative manner

(b). These expert judges were not randomly chosen and analyzed the same set of test items, but were selected based on specific criteria (see content validity section). This was done for a *Task Difficulty Score* (Diff-Sc; low, medium, high difficulty) that reflected the impressions of the difficulty among test items, for a *DigComp Relation Score* (DigCR-Sc; numerical, 0 to 100) as stated above, and for a *Task Complexity Score* (Comp-Sc; numerical, from 1 to 7, in accordance with DigComp-related complexity) reflecting the complexity of the items given a pool of options. So, it was expected this MDAT construct would show: (1) low to moderate complexity, (2) low difficulty, and (3) a relation score above 50 for each major DigComp dimensions.

For the ICC calculation it was applied a two-way mixed effects model, studying consistency, for multiple raters and measurements, to establish measures of absolute agreement (Shrout & Fleiss, 1979) in accordance with nomenclature conventions (McGraw & Wong, 1996). The interpretation of ICC values followed standard parameters (Bobak et al., 2018; Cicchetti, 1994; Koo & Li, 2016).

SmartPLS version 4.0.9.4 was used to conduct factor analysis statistics to identify latent variables and structural equation modeling. Reliability was assessed using Cronbach's alpha coefficient ($\alpha$), with values $\geq 0.70$ taken to indicate good internal consistency. Cronbach's $\alpha$ was calculated for the whole scale, and distinct subsets of items. Additionally, bootstrap methodology (10.000 bootstraps) was used in SEM models to derive standard errors and p-values, to provide estimates of the parameters. By adopting this statistical methodology, a more detailed representation of the data was expected.

During the planning phase of the MDAT research, a structured study size statistical evaluation was implemented. The analysis after statistical consulting employed both three-group ANOVA and two-group pairwise comparisons, exploring different age groups performance, utilizing previous results in two different populations as a reference for mobile digital device proficiency (Moret-Tatay et al., 2019; Roque & Boot, 2018). Methodological decisions, including statistical parameters, were done with G*Power, adhering to an alpha of 0.05 and targeting a power of 0.8 (Faul et al., 2009). This approach provided a better understanding of test performance, aligning with best practices with adequate *a priori* sampling.

**Criterion-related statistical validity analysis**

The concurrent validity of the MDAT test was analyzed by correlating its scores with measures such as the amount of time they interact with mobile devices. Pearson's correlation coefficients were calculated to quantify the strength of the relationship between the MDAT and other measures for normal distributed data. Non-parametric correlation methods were applied for non-normal distributions. The association between variables in the dataset paired observations (n = 101) was done with Spearman's rank correlation coefficient ($\rho$) and given the possibility presence of tied ranks also with Kendall's tau ($\tau$), which served to ascertaining the strength and directionality of association, integrated to the assessment of correlation, and can have interpretable results and be more efficient than Pearson method in some situations (van den Heuvel & Zhan, 2022). Calculated *p-values* were expressed and *p-value* threshold of < 0.05 was used to infer statistical significance. In line with published guidelines, we interpreted correlation coefficients ($x$) as follows: small ($x \leq 0.1$), medium ($x \leq 0.3$), and large ($x \leq 0.5$) (Cohen, 1988). High or moderate positive correlations were thought to provide robust support for the concurrent validity of the MDAT scores. Positive significant correlations between mobile usage time and daily frequency were expected to be positively correlated with the best fit MDAT model scores.

**Data management**

For a more rigorous data collection, a standardized protocol was followed and data were collected in consistently across the study by using REDCap electronic data capture tool hosted at Paulista Medical School, at the Federal University of São Paulo (Harris et al., 2009). REDCap (Research Electronic Data Capture) is a secure web-based software platform designed to support data capture for research studies by providing (1) an intuitive interface for validated data capture; (2) audit trails to track data manipulation and export procedures; (3) automated export procedures for continuous data downloads to common statistical packages; and (4) procedures for data integration and interoperability with external sources (Harris et al., 2019). Once collected, the data underwent quality control and preprocessing. For missing data, a case-wise deletion was applied to instances where more than 5% of relevant variables were missing, giving rise to a studied population of 101 individuals of 105 attending tested subjects. The missing data was considered missing at random, and this assumption was deemed reasonable given the context of our study and the low rate of missing data. Data were organized and structured for statistical analysis using R software or JASP.

# Results

## Sample description

  The collected sample allowed the research to analyze the modeled interplays between the sociodemographic factors and digital mobile device usage. The sample was considered diverse and was composed by adults (n=101) majorly females (76.2%; n=77), smaller number of males (23.8%; n=24), representing multiple ethnicities: White (43.6%; n=44), Black and Pardo(a) (53.5%; n=54). Regarding education, the average of study years was 12.0 (SD=4.3), with participants classified into three categories: less than 6 years of schooling (8.9%; n=9), 6 to 12 years of schooling (39.6%; n=40), and more than 12 years of schooling (51.5%; n=52). Participants included employed individuals (45.5%; n=46), unemployed (9.0%; n=9), and retirees (25.7%; n=26). The average per capita income was R\$ 2,250.00 (equivalent to \$444.90 USD on 31st May 2023, not inflation-adjusted). The sample had a low to middle income population. We also assessed participants' health risk factors, such as their engagement in physical activity, smoking status (past/current smoker), alcohol use (weekly use), and obesity. See Table 1 for further information.

**Table 1.** Sample characteristics.

|  | Total sample (n=101) | 20 to 39 years (n=27) | 40 to 59 years (n=42) | 60 to 79 years (n=32) | p-value |
|---|---|---|---|---|---|
| Sex |  |  |  |  | 0.191 |
| Female | 77 (76.2%) | 23 (85.2%) | 33 (78.6%) | 21 (65.6%) | - |
| Male | 24 (23.8%) | 4 (14.8%) | 9 (21.4%) | 11 (34.4%) | - |
| Brazilian ethnicity |  |  |  |  | 0.377 |
| White | 44 (43.6%) | 9 (33.3%) | 15 (35.7%) | 20 (62.5%) | - |
| Black | 19 (18.8%) | 4 (14.8%) | 11 (26.2%) | 4 (12.5%) | - |
| Pardo(a) | 35 (34.7%) | 12 (44.4%) | 16 (38.1%) | 7 (21.9%) | - |
| Marital status |  |  |  |  | < 0.01 [§] |
| Never married | 26 (25.7%) | 19 (70.4%) | 3 (7.1%) | 4 (12.5%) | - |
| *Currently married* | 30 (29.7%) | 1 (3.7%) | 19 (45.2%) | 10 (31.3%) | - |
| *Live with partner* | 14 (13.9%) | 1 (3.7%) | 9 (21.4%) | 4 (12.5%) | - |
| Study (years) | | | | | < 0.01 |
| *Mean (SD)* | 12.0 (4.34) | 14.5 (2.65) | 12.6 (3.50) | 9.13 (4.91) | |
| Schooling years | | | | | 0.027 |
| *Less than 6 years* | 9 (8.9%) | 0 (0%) | 0 (0%) | 9 (28.1%) | - |
| *From 6 to 12 years* | 40 (39.6%) | 6 (22.2%) | 18 (42.9%) | 16 (50.0%) | - |
| *More than 12 years* | 52 (51.5%) | 21 (77.8%) | 24 (57.1%) | 7 (21.9%) | - |
| Work status | | | | | < 0.01 [§] |
| *Has paid job* | 30 (29.7%) | 7 (25.9%) | 18 (42.9%) | 5 (15.6%) | - |
| *Self-employed* | 16 (15.8%) | 4 (14.8%) | 9 (21.4%) | 3 (9.4%) | - |
| *Retired* | 26 (25.7%) | 0 (0%) | 5 (11.9%) | 21 (65.6%) | - |
| *Per capita* income | | | | | |
| *Mean R$ (SD)* | 2,250 (2,180) | 2,080 (2,990) | 1,880 (1,280) | 2,890 (2,240) | 0.053 |
| Health risk factors | | | | | |
| *Physically active* | 50 (49.5%) | 12 (44.4%) | 23 (54.8%) | 15 (46.9%) | 0.661 |
| *Past/current smoker* | 35 (34.7%) | 5 (18.5%) | 14 (33.3%) | 16 (50.0%) | 0.039 |
| *Weekly alcohol usage* | 42 (41.6%) | 10 (37.0%) | 17 (40.5%) | 15 (46.9%) | 0.734 |
| *Current obesity* | 24 (23.8%) | 2 (7.4%) | 14 (33.3%) | 8 (25.0%) | 0.046 |

Variable distributions reported as number (N) and per cent frequency (%), unless otherwise specified. Numeric variables are reported with mean values. The self-reported *pardo* description refers to an ethnic and skin colour category used by the Brazilian Institute of Geography and Statistics in the Brazilian census. The term is complex, used to refer to Brazilians of mixed ethnic ancestries (Oliveira, 2004). This sample table was built following recommended formatting, with *table1* R Package (Hayes-Larson et al., 2019; Rich, 2023). For checking some differences among the age groups (exact p-values shown), an ANOVA was used for numerical variables, and a Chi-Square for categorical variables. A Monte Carlo simulation ($10^3$ replications) was done to estimate the *p-values* for work status and marital status (§) variables due to constraints with traditional tests computations. Other ethnicity, marital status, work status beyond the described are not shown. Abbreviations: SD, standard deviation; R$, Brazilian real currency.

**MDAT content validity analyses**

Using the CVI and unanimous opinion (UA) calculated scores among the experts panel, it was possible to observe that the MDAT achieved good or very good content validity, with S-CVI-Av of 0.982 and S-CVI/UA of 0.875, respectively, taking into account the 4 aspects investigated – (a) clarity of language, (b) inferred understanding (by the target population), (c) item accuracy (practical relevance for measuring this task in the real world), and (d) comprehensiveness (in relation to the theoretical construct of the test) – all reaching satisfactory quantitative scores. The item the lowest score was Task 2 and impacted the possible comprehension of the participants (individual details not shown). All suggestions of text modification that received more than 40% of suggestions (or 3 suggestions out of 7 experts) were accepted. The results are shown in Table 2.

**Table 2.** Items validity analysis.

|  | S-CVI/Av | S-CVI/UA |
|---|---|---|
| a. Language Clarity | 0.976 | 0.833 |
| b. Inferred Comprehension | 0.952 | 0.667 |
| c. Item Accuracy | 1.000 | 1.000 |
| d. Comprehensiveness | 1.000 | 1.000 |
| Average CVI analysis | 0.982 | 0.875 |

Note: Averaged scaled Content Validity Index (S-CVI/Av) were considered very good for (a) language clarity and (b) comprehension (i.e., how the experts think the item would be properly understood by participants), although lower experts' unanimous agreement (S-CVI/UA). Better scores were measured for (c) item accuracy (supposed items capability of real-world data representation) and (d) comprehensiveness (for the items inside the test construct).

As for the study of construct content in relation to correlated DigComp tasks, it was observed that different tasks of the MDAT test, in the opinion of the panel of experts, have different profiles of relationship with the Competence Areas of DigComp 2.2. The questions asked to them ware, for example: *"How does the Task 1 ('unlocking the device') 'relate' to this kind of competence dimension described in DigComp?"* The answer was given on a visual analog scale (VAS), where on the left, there was the minimum ability (zero points), and on the right, the maximum (100 points), called "DigComp Relation Score" (DigC-Sc). As expected, it was observed that each basic MDAT task seemed more related to a specific DigComp Competence area, in accordance with the expert panel opinion: *Task 1* with "Security" competence are, *Task 2* with "Communication and Collaboration", *Task 3* with "Content Creation", and *Task 4* with "Problem Solving". See Table 3 for details.

**Table 3.** Digital competence areas and MDAT tasks comparison.

| DigComp 2.2 Competence Areas | MDAT Tasks (DigComp Relation score) | | | |
| --- | --- | --- | --- | --- |
| | Task 1 *(unlock)* | Task 2 *(message)* | Task 3 *(photo)* | Task 4 *(turn off)* |
| 1. Information and data literacy | 24 | 64 | 54 | 40 |
| 2. Communication and collaboration | 16 | **96** | 84 | 23 |
| 3. Content creation | 14 | 82 | **92** | 13 |
| 4. Security | **90** | 71 | 68 | 59 |
| 5. Problem solving | 74 | 83 | 77 | **69** |

The table show comparisons between DigComp digital competence areas and the corresponding MDAT tasks (DigComp Relation Score, *DigC-Sc*). The numerical values, attributed as a 0 to 100 scores by the 7 experts with a visual analog scale (VAS), represent the average they attributed on each task (n=7, deviation not shown). The competence areas are shown, numbered 1 to 5. The tasks are labeled 1 to 4, and highest attributed *DigC-Sc* values are highlighted in bold.

**Rater agreement for content parameters**

An assessment of the Intraclass Correlation Coefficient (ICC) was conducted for three distinct content data-based calculated scores, named Diff-Sc, Comp-Sc, and DigC-Sc, with the intent of checking how uniform the content evaluation was. The inter-rater reliability was calculated among the 7 judges, specific to the raters utilized in this study, without broader generalization. The Diff-Sc, "difficulty attributed score", yielded the relatively lower ICC point estimate, with value of 0.238, demonstrating a significant variability among raters (95% CI = 0.104 to 0.439), and suggesting individual inner-group discrepancy in 'difficulty' interpretation. The Comp-Sc, "complexity attributed score", showed the higher degree of homogeneity among the raters with an ICC point estimate value of 0.839 (95% CI = 0.740 to 0.915), indicative of good to excellent inter-rater reliability. This data showed that the raters agreed that the MDAT tasks here described were easy, as planned. Lastly, the DigC-Sc, DigComp relation score, quantified on the Visual Analogue Scale (VAS), attributing quantitative "relational" scores from 0 to 100 for each task (see Table 3), showed a moderate

ICC point estimate of 0.556 (95% CI = 0.386 to 0.732), suggesting a fair/moderate degree of concordance.

**Table 4.** Experts panel evaluations and agreement of the complete test set.

| Evaluation Scores | Descriptives | | | Interrater agreement | |
|---|---|---|---|---|---|
| | Range | Average | Median (IQR) | ICC estimate (95% CI) | ICC interpretation |
| Difficulty | 1 to 3 | 1.85 | 2.0 (1.0 - 2.0) | .458 (-.004 to .754) | Low |
| DigComp Relation | 0 to 100 | 59.5 | 70 (30 - 90) | .907 (.828 to .958) | Good/Excellent |
| Complexity | 1 to 8 | 3.46 | 3.0 (2.2 - 5.0) | .957 (.920 to .981) | Good/Excellent |

Median, average, and interquartile range (IQR) for attributed/evaluated *Tasks Difficulty* (ordinal, low/medium/high, range: 1 to 3), *DigComp Relation* (numerical, range: 0 to 100, visual analog scale), and *Tasks Complexity* (ordinal, lower to higher, range: 1 to 7). Observe that tasks difficulty and complexity were considered low to medium valued, as expected. DigComp relation above 50 thresholds on average (compare with Table 3 data). Point estimates for Intraclass Correlation Coefficient (ICC) for 7 raters are also shown. Each measurement was rated by the same fixed set of raters and tests (raters of interest) and the results show averaged ratings (team consistency). ICC (3, k), in accordance with conventions. Diff-Sc, difficulty attributed score, was ranked for the 4 MDAT tasks. Comp-Sc, tasks complexity attributed score, in accordance with DigComp proficiency level. DigC-Sc, the DigComp relation score. CI, confidence interval. Colours attributed are consistent with its values.

**MDAT model selection and evaluation with PLS-SEM**

The MDAT model selection followed a stepwise approach to find the best performing model. Each model run, as stated, was tested based on pure data analytics parameters, assuming that they are correlated at least with moderate/high positive correlation scoring (measured by PLS-SEM adjusted $R^2$) with the comparative standard (contact/interaction with mobile devices), which was consistently achieved with all models tested. As initially assumed, the best model would be the one that (a) correlated best with usage time and had (b) optimal internal consistency. Variables used to build the models are detailed in Table 5.

**Table 5.** Comparison of model specifications.

| Model | Usage time variables | MDAT proficiency score variables included | Moderator variables | Control variables |
|-------|----------------------|-------------------------------------------|---------------------|-------------------|
| Model 1 | M1, M2, C0 | Task 1, Task 2, Task 3, Task 4, mdat_time, mdat_rept | None | None |
| Model 2 | M1, M2 | Task 1, Task 2, Task 3 | None | None |
| Model 3 | M1, M2 | Task 1, Task2, Task 3, mdat_rept | None | A2, A3 |
| Model 4 | M1, M2 | Task 1, Task 2, Task3, mdat_rept | A1 | A2, A3 |
| Model 5 | M1, M2 | Task 2, Task 4, mdat_time, mdat_rept | A1 | A2, A3 |

Differences in model specifications across the runs (*Model 1 to 5*) are shown. The last model (*Model 5*) had the best measured performance. Each model had different variables set for latent variables (*usage_time, proficiency_score*), as well as the presence of a moderator and control variables. Variables codenames: *M1*, mobile daily usage (ordinal); *M2*, usage time in years (ordinal); *C0*, computer daily usage (yes/no); *Task 1 to 4* (yes/no); *mdat_time* (continuous), total time in seconds to complete the task; *mdat_rept* (integer), number of needed instructions reads to the participant; *A1*, study years (continuous); *A2*, age in years (continuous); *A3*, family income *per capita* (continuous).

Considering that two latent variables, MDAT proficiency and mobile device interaction are positively correlated, it was assumed that the $1^{st}$ model would include all of them. MDAT proficiency (*proficiency_model*) was regressed on mobile device interaction (*usage_time_model*), assuming a directional relationship between the two latent variables. Unexpectedly, weekly personal computer usage did not add much value to the model, based on correlation with the predicted score. Usage was operationalized using three ordinal variables containing contact time with mobile technology (see Table 4), while MDAT proficiency was operationalized using 6 variables: the completed MDAT Tasks 1 to 4 (completed yes/no, binary), time to fully complete the test (seconds, numerical, reversed order) and times instructions were read (times, numerical, reversed order) for that participant (*mdat_repet* and *mdat_time*).

Based on the results of the 1st model run, MDAT *Task 4*, *mdat_repet* and *mdat_time*, and *C0* variables were initially removed due to insufficient reliability and convergent validity. The 2nd model run showed improved results compared to *Model 1*. In the 3rd model run, age (*A2*) and income (*A3*) were added. Results were equivalent to those at the *Model 2* run, and no major influence by these variables was perceived, in opposition to the previous assumption that they could have major impact on measured MDAT proficiency. The 4th model run introduced the schooling years as a moderator variable (*A1*). The results showed an inverse correlation between *A1* and MDAT *proficiency_model*, suggesting moderating effect of *A1* on the relationship between *usage_time_model* and MDAT *proficiency_model*. At last construct tested (*Model 5*), age and income were also included as control variables for MDAT *proficiency_model*. The inclusion of these variables led to a less significant correlation between *A1* and *proficiency_model*. However, results also showed a moderate positive correlation between *A1* and *A2*, while *A3* contributed minimally to the explanatory power.

It is important to report that before MDAT model construction with path models using SEM technique, there were created some dichotomized data analysis of the MDAT Tasks. Although they were not ideal for reasons highlighted before (see Methods), there may be clinical utility to create scores that are easily calculated. Five models, *D1* to *D5*, were additionally developed and tested against the MDAT *Model 5* using PLS-SEM. The best correlating was the *D4* model. This dichotomized data analysis was not included here.

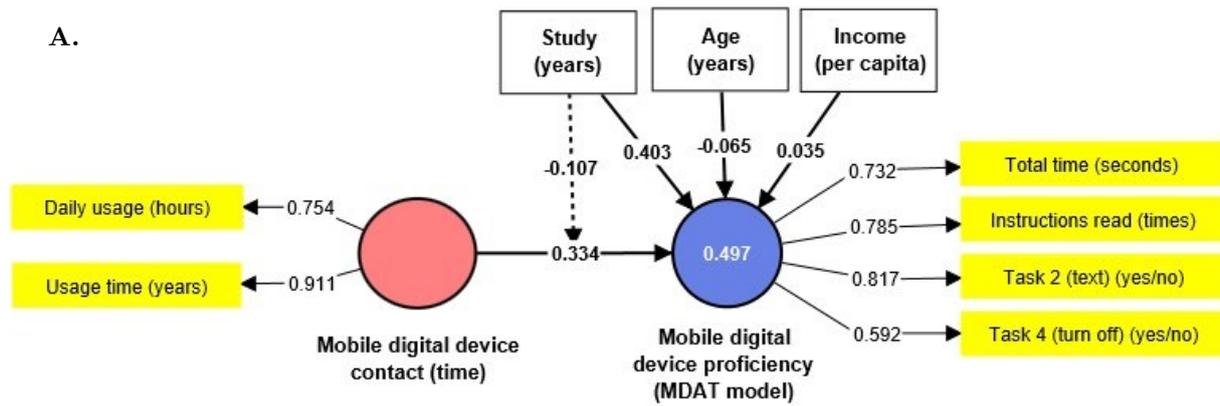

**A.**

| **B.** Model latent variables | MDAT proficiency | Device usage | Criteria | |
|---|---|---|---|---|
| MDAT proficiency | **.779** | - | Diagonal > out of diagonal range | Discriminant validity |
| Device usage | **.523** | **.828** | | |
| **C.** Model parameters (additional) | | | | |
| Cronbach's alpha (α) | **.712** | .587 | ≥ 0.7 | Reliability |
| Composite reliability (rho_A) | **.721** | .671 | ≥ 0.7 | Reliability |
| Composite reliability (rho_C) | **.824** | .821 | ≥ 0.7 | Reliability |
| Average variance extracted (AVE) | **.543** | .699 | ≥ 0.5 | Convergent validity |

**Figure 2.** MDAT model characteristics. **(A)** MDAT, graphical representation for the PLS-SEM path model, for *Model 5*. The light (red and blue) circle shapes represent the latent constructs ("device contact time" and "device proficiency", respectively). The outer model (arrows' direction: latent → measured variables, as shown) is represented with outer loadings: all are significant (p < 0.001). Outer loadings are above 0.7 except for Task 4 (0.59). It is showing good convergent validity and reliability at item level. The inner model (or structural model) is represented by path coefficients (exogenous variables → endogenous variable). The traced line represents the moderating effect of the study years in the relationship between MDAT and contact time. The R² value for MDAT is 49%. All paths are significant (p < 0.001), except for (i) the moderation (-0.107, p > 0.05), (ii) age → MDAT (-0.065, p > 0.2), and (iii) income → MDAT (0.035, p > 0.2). **(B)** Table showing validity criteria, with reliability, discriminant, and convergent validity analysis. Note. Diagonal values = root square of AVE. Out of diagonals = correlation between latent variables. To evaluate them, established guidelines were followed (Hair et al., 2021).

## Concurrent and nomological validity and motor skill correlation

To test for additional validity measures and to re-examine association between out-of-model variables, we examined the correlations between the scaled MDAT scores (calculated scores, ranging from 0 to 100). Figure 3 summarizes these results. As expected, there was a strongly negative correlation between MDAT scores and age (years, -0.49, *p* < 0.05), and positive with formal schooling (years, 0.59, $p < 0.05$). Confidence intervals (CI) for these correlations were obtained after bootstrapping with $10^3$ resamples (see Figure 3A).

Younger participants had not normally distributed MDAT scores (Shapiro-Wilk *p-value* < 0.01). So, non-parametric Kruskal-Wallis's test was then used to check differences among the 3 groups: younger (20-39 years), middle-aged (40-59 years), and older (60-79 years) adults. The test results showed significant variance among MDAT scores between the groups ($H = 22.9$, df = 2, $p < 0.01$). Post-hoc analyses were conducted to evaluate between-group differences (Dunn's and Holm's methods). It showed that the younger group (mean MDAT = 86.6, SD = 19.3, N = 27), differed ($p_{Holm} = 0.03$) from the middle-aged group (mean MDAT = 74.8, SD = 28.3, N = 42). The younger group diverged significantly ($p < 0.01$) from the older group (mean MDAT = 58.0, SD = 26.6, N = 32). The middle-aged *versus* older group score difference was also significant ($p_{Holm} = 0.01$). Figure 3C shows MDAT scores across age groups. Different mobile operating system usage was also examined (Figure 3F).

Functional auto-perception measured by WHO-DAS 2.0 12 items questionnaire was not different among groups (grouped ANOVA statistics, non-parametric, n = 101, $p < 0.01$). The average WHO-DAS 2.0 score was 15.09 points (SD 10.73, ranging from 0.0 to 43.7). Quality of life scores measured by EQ-5D-5L (Figure 3D) psychometric instrument were also non different among groups (mean = 0.917, SD = 0.109, range = 0.476 to 1.000). This ensures the scores have lower influence from health measures, and maybe an independent psychometric property to study (see Figure 3B).

Motor task (dexterity) tested with 9-HPT was assessed for both hands. The average test time for right hand (9-HPT Right) was 23.26 seconds (SD = 3.21, IC SD 95% = 2.72 to 3.66 sec.), and 24.69 seconds (SD = 3.65, IC SD 95% = 3.17 to 4.03 sec.) for left hand (9-HPT Left). There was no major influence of dexterity on MDAT scores (PLS-SEM factor loading and correlation analysis, not shown), but there was a negative correlation coefficient (-0.22, p = 0.027, CI 95% -0.41 to -0.02) between MDAT scores and 9-HPT Right time, meaning that if a participant had a worse time taken for this dexterity test with his right hand (more seconds to finish the task), the worse the MDAT score was (Figure 3A).

| Correlates of MDAT scores (n = 101) | | Spearman's ρ | p-value | 95% CI | |
|---|---|---|---|---|---|
| Demographic | Age (years) | **-.490** *** | .001 | -0.641 | -0.320 |
| | Study (years) | **.594** *** | .001 | 0.454 | 0.695 |
| | Income(BRL) | .153 | .125 | -0.026 | 0.321 |
| Psychometric | WHO-DAS 2.0 | -.053 | .601 | -0.244 | 0.146 |
| | EQ-5D-5L | .171 | .088 | -0.012 | 0.353 |
| | 9-HPT Right | **-.221** * | .027 | -0.417 | -0.022 |
| | 9-HPT Left | -.135 | .178 | -0.331 | 0.065 |
| Latent variable | Usage (contact) score | **.466** *** | .001 | 0.286 | 0.622 |

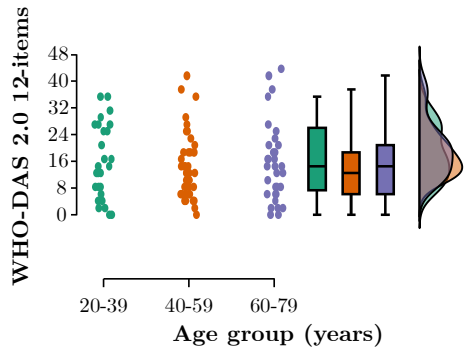

**B.**

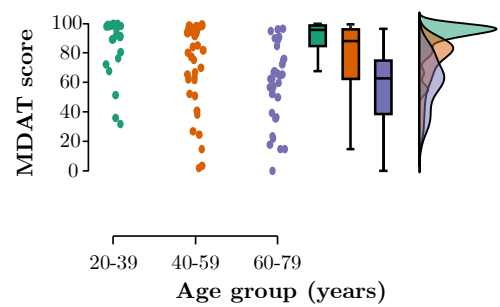

**C.**

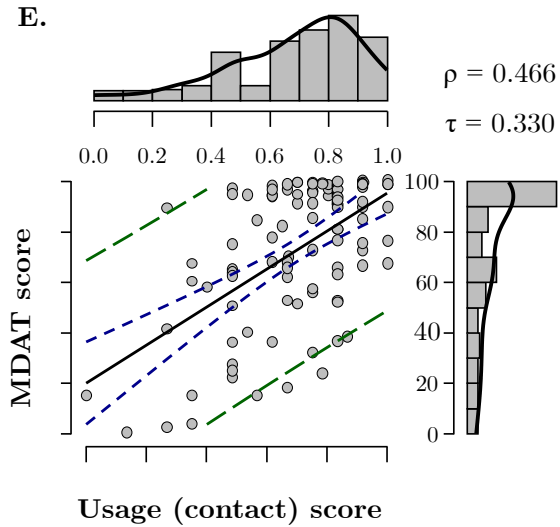

**E.**

ρ = 0.466

τ = 0.330

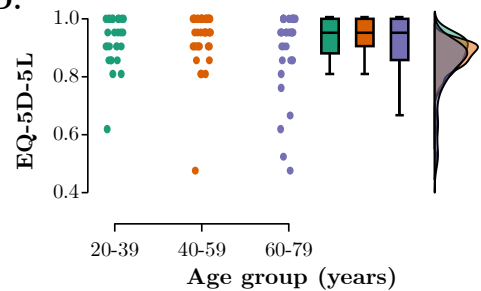

**D.**

**F. Mobile operating system use, n (%)**

| Age | Android OS | iOS |
|---|---|---|
| ≤ 50 years | 29 (82.8%) | 5 (14.2%) |
| > 50 years | 21 (87.5%) | 3 (12.5%) |

**Figure 3.** MDAT comparisons. **A.** Correlates for MDAT scores, with Spearman coefficients (ρ) and confidence intervals (95% CI) for demographic, psychometric and mobile device usage (contact) latent variable. Significance levels are highlighted with asterixis. Table colors indicate correlation strength (red for negative and blue for positive ρ values). **B, C, D.** Grouped analysis by age groups (20-39, 40-59, 60-79 years-old) for auto-perceived global functionality (WHO-DAS 2.0), quality of life or global perceived health (EQ-5D-5L), and MDAT mobile proficiency modeled scores. Note that there is a significant difference for the age groups (ANOVA, $p = 0.01$) for MDAT (generational gap), but not for WHO-DAS or EQ-5D-5L in this sample. **E.** Dot plot graph (with densities), MDAT scores *versus* Usage (contact) score (latent construct). Spearman's rho (ρ) and Kendall's tau (τ) correlation coefficients are shown. **F.** Table for older and younger participants who reported (%) their most used OS (Google Android OS, or Apple iOS) by themselves, on mobile devices (N=59). Chi-Squared test showed no difference between groups ($\chi^2 = 0.755$, DF = 2, N = 59, $p = 0.685$), i.e., no significant association between the two variables (less/equal or more than 50 years old) in this contingency table. Similar result was obtained for a 3-age-group analysis ($p = 0.324$).

# Discussion

The MDAT was developed as an open-source psychometric testing framework, to meet the need for a simulated real-world performance-based to measure mobile device proficiency aimed at cognitive health. The demand was identified by our group, health care professionals with expertise in Behavioral and Cognitive Neurology. Optimal application of MDAT could improve mobile device proficiency stratification between individual with low or high abilities and checking for digital divide and proficiency gaps across different groups. This can lead to proper methodological evaluation standards and conscious detection of low ability individuals that may not be suitable for mobile digital health interventions, especially considering the additional DigComp correlations for validity and reliability described at the study.

There are some limitations to be considered. To the best of the group knowledge there is no other known performance-based tests to compare the MDAT for concurrent validity and the group is not aware of a gold standard. There was some similar research but not with these specific intent (Petrovčič et al., 2019). The MDAT framework may be oversimplistic, as it does not cover the whole *DigiComp* competences framework construction list. Although item selection was based on theoretical and empirical rationale, systematic review show most cited competency areas on reference frameworks are problem solving (n=10), safety (n=7), information processing (n=5) and content creation (n=5), communication (n=2), matching with the MDAT item selection described here (Tinmaz et al., 2022). Besides carefully done evaluations, the final test set here described is really a particular way to interpret data. The suggestion is that the same could be done for multiple devices. The MDAT test framework could be difficult to reproduce, so it is advised to consider testing this framework method with different devices and OS, to see how scalable it could be.

In-depth correlations with cognitive screening tests would give more insights and data into how this can be used. Additionally, there is a floor effect of the WHO-DAS utilized here as a global functioning evaluation, so it is a problem to test otherwise "normal" subjects (Saltychev et al., 2021). More cognitive functional tests could add up. The test framework is not yet tested with other data sets with similar data (Rahman et al., 2022), and it is really an one-site study, so data may not represent other populations, even for specialized centers. Another methodological problem is related to usage. PLS-SEM-derived MDAT scores are not readily calculable for a clinical purpose, so one could use dichotomization for simpler calculation, but with the hurdle of poorer data representation. An option to this could be to standardize all the measures (z=score) and use the average of the items as the score for each construct, in this case: MDAT and mobile digital device contact (time). Ethical questions regarding digital interventions to determine cognitive difficulties are relevant but not fully adopted (Ford et al., 2023; Ursin et al., 2021). Other improvements would include two measure points of this same construct and have retest data. Time constrains for the pilot project, lack of resources, and difficulty around healthy volunteer recruitment were impeding factors. There are plans to have validated instruments to compare with the MDAT.

There are other suggestions for future MDAT research framework improvement. (1) As *Task 4* had a loading = 0.59, the item could be revised so that this could be increased to 0.7, which is considered ideal. (2) As also identified, there was no task included that had a strong correlation with the DigiComp competence "1. Information and Data Literacy" (Table 3). It would be possible to add another task that has a better correlation, or perhaps revise *Task 4* itself so that we have a stronger correlation some competence (average > 80), but with the problem of increasing the testing time. (3) There was a low ICC score in the expert panel's assessment of the average of the tasks analyzed in relation to their "Difficulty", although "Complexity" was within expectations (Table 4). There is central tendency bias and reduced sensitivity/specificity for fewer discriminants. It is possible to adjust the tasks to obtain an average difficulty closer to low/medium scores (i.e., equal to 1) as it was expected, or redo an expert panel (independent), but using a Likert-type coding that can better detect (with better discrimination) these differences. Using a 5-level Likert scale instead of 3 levels (range: low, medium, high) is shown to have better discriminant capabilities (Preston & Colman, 2000).

## Conclusions

The aim of this research was to develop an instrument to assess proficiency in the use of digital mobile devices that was easily executable but valid and based on performance. Before applying it to cognitively healthy volunteers, the items of the instrument were carefully chosen, and according to evaluations with an interdisciplinary panel of experts, refined for its content and construct. The analysis of the 101 volunteers tested with the instrument obtained a good average completion time (3.56 minutes) and an average MDAT proficiency score of 72.6 points (standard deviation = 27.7 points, on a scale calculated ranging from 0 to 100 points). It was possible for the instrument to differentiate between the three main age groups in terms of their proficiency in using digital mobile devices, in the same way as in equivalent research. In addition to discriminatory capabilities, the test appears not to be as influenced by self-reported global functionality measured with the WHO-DAS 2.0. Although an adapted setting is necessary to carry out this type of assessment, it has a low cost, depends on a simple smartphone-type mobile phone, and is quite feasible even in environments with few resources. Motor dexterity and thumb clicking speed on both hands had no impact on the test, showing that motor function may not be relevant in this context.

The result is a fast, open-source testing framework capable of providing user-centered information that can be included in cognitive health research. Thus, we believe that the MDAT can be present in more research sites and be compared with other psychometric instruments. The authors hope for greater inclusion of proficiency measures in digital health research, something that is still incipient. The final step is to have end-user and patient-centered strategies adding new insights and facilitating at-risk patients with mild cognitive impairment before an impactful cognitive dysfunctions and dementia affects their health.

## Data availability

Due to the nature of the research, participants of this study did not agree for their data to be shared publicly or to third party outside the research team, so supporting granular data is not available. Requests for data will be considered by the DigiTAU study research team based on scientific priorities and overlapping interests. A deidentified database is planned to be published under ethical research practices after studies completion. Requests to access these datasets should be directed to Alan Cronemberger Andrade (alan.andrade@unifesp.br).

## Authorship contributions

All authors were involved in reviewing the literature, writing, and commenting on the manuscript. Study concept and design (ACA, PHFB); MDAT framework test (ACA); acquisition of data (ACA); analysis and interpretation of data (ACA, ACB, DB, WB, PHFB); drafting of the manuscript (ACA, ACB, PHFB); critical revision of the manuscript for important intellectual content (ACA, PHFB); statistical expertise (DB); obtained funding (PHFB); administrative, technical, and material support (ACA, PHFB); study supervision (PHFB).

**License**

This test framework is free and explicitly designed for academic, educational, health, and research applications at accredited institutions. It may be freely used, modified, or distributed for any purpose, including commercial applications. Any modifications, derivative works, or adapted versions must also be made freely available for non-commercial purposes, particularly for academic and research use. In the case of commercial applications, the derivative works must be distributed under identical terms, making them freely available for academic and research use. Users are not required to obtain prior authorization from the authors for any usage. When publishing or distributing the test or any derivative works, users must provide proper attribution to the original authors. While it is not a legal obligation, users who intend to utilize this test for commercial applications are encouraged to notify the corresponding author. Instructions and materials for this test are freely accessible in both Portuguese and English at the Open Science Framework (OSF), in alignment with open science principles. Collaborative contributions to the refinement and expansion of this test are encouraged.

**ORCID**

Alan Cronemberger Andrade MD MSc                    https://orcid.org/0000-0003-1645-0500

Diógenes de Souza Bido MSc PhD                          https://orcid.org/0000-0002-8525-5218

Ana Carolina Bottura de Barros MSc PhD            https://orcid.org/0000-0002-1879-3702

Walter Richard Boot PhD                                       https://orcid.org/0000-0003-1047-5467

Paulo Henrique Ferreira Bertolucci MSc PhD MD   https://orcid.org/0000-0001-7902-7502

**Disclosure of conflict of interest**

The authors declare no competing interests.

**Acknowledgements**

This study at the Postgraduate Program in Neurology and Neuroscience was financed in part by the *Coordenação de Aperfeiçoamento de Pessoal de Nível Superior* (CAPES), Finance Code 001, Brazilian Ministry of Education, Brazil. We are immensely grateful to all study participants and all those candidates who show interest in participating, the helpful employees at Hospital São Paulo and the Federal University of São Paulo, and colleagues from the Behavioral Neurology Sector, Federal University of São Paulo. A special thank you also goes to Dr. Walter Teixeira Lima Junior, and the Artificial Cognitive Systems Research Group, for the words of encouragement and support.

# Supplementary Material

## MDAT Portuguese Language Instructions

## TESTE DE HABILIDADES DIGITAIS (THD)
### INSTRUÇÕES PARA REALIÇÃO DESTE TESTE

### LEITURA DE INSTRUÇÕES ORAIS AOS RESPONDENTES

"Faremos um breve teste para avaliar suas habilidades de uso de aparelhos móveis digitais. Você usará este aparelho preto que está a sua frente para executar algumas tarefas que vamos ler para você. Quanto mais rápido realizar o teste melhor poderá ser o seu resultado. Sabemos que este não é o seu aparelho. E, não se preocupe se não conseguir fazer alguma tarefa. Cada instrução deve ser realizada na exata ordem em que foram lidas, e só comece depois que terminar de ler as tarefas. Só iremos ler as instruções uma vez antes do teste. Se precisar, poderemos ler de novo as instruções, mas cada nova leitura será anotada, e você perderá "pontos". Se não conseguir executar e quiser pular uma tarefa, podemos fazer isso, e só pedir. Mas, é pior não conseguir fazer uma tarefa que pedir para repetir uma instrução. As instruções são 4 (quatro), são 4 tarefas:

(1) Por favor, <u>desbloqueie o celular</u>. Isso pode ser feito pela tela, pelo botão que está localizado à direita do aparelho, ou pelo botão ao centro. O desbloqueio deste aparelho específico é por um código numérico. Para desbloquear o celular o código é 2584. Após concluir a tarefa, daremos aviso que concluiu.

(2) <u>Escreva uma mensagem</u>. Na tela, você encontrará três aplicativos usados para as pessoas se comunicarem. Identifique aquele que você conseguirá usar para enviar uma mensagem ao contato "Pedro". A mensagem deve conter: "Pedro, vou na farmácia ao meio-dia.". Após concluir a tarefa, clique em enviar, e daremos aviso que concluiu.

(3) Procure um aplicativo para fotografia. <u>Tire uma foto</u> centralizada da bola laranja que está ali (apontando) atrás de você. Após concluir a tarefa, daremos aviso que concluiu.

(4) Após terminar <u>desligue o celular</u>. O teste estará concluído após desligar o aparelho, apertando e segurando o botão localizado a direta do aparelho, e clicando em desligar, não somente bloqueando a tela. Tem que desligar o aparelho."

### INSTRUÇÕES AOS APLICADORES OU EXAMINADORES

(1) Após ler estas instruções, entregue o dispositivo para a pessoa que está sendo testada. (2) Após esta leitura, entregue-o ele/a ela o mais rápido possível, e comece a contar o tempo. (3) Lembre-se de avisar em voz alta (ou áudio automatizado de alto volume) o término/conclusão de cada tarefa. Diga apenas: "Tarefa concluída". (4) Não se esqueça de anotar o tempo total e de cada tarefa (a) e a quantidade de leitura de instruções (b), e quais tarefas foram realizadas (concluída, sim ou não) (c). (5) Automatize a tomada de tempo o máximo que puder a fim de evitar erros e grave o teste, com áudio e vídeo, sempre que possível.

# Supplementary Material

**MDAT English Language
Instructions**

# MOBILE DEVIDCE ABILITIES TEST (MDAT)
## TEST EXECUTION GUIDELINES

### ORAL INSTRUCTIONS FOR RESPONDENTS

"We will conduct a brief test to assess your skills in using digital mobile devices. You will use this black device in front of you to carry out a few tasks that we will read out to you. The quicker you complete the test, the better your score could be. We are aware this is not your personal device, and don't worry if you cannot accomplish a task. Each instruction should be carried out in the exact order it was read, and you should only start after we finish reading the tasks. We will only read the instructions once before the test. If needed, we can read them again, but each new reading will be noted, and you will lose "points". If you can't perform a task and want to skip it, we can do that, just ask. But it's worse to not be able to accomplish a task than asking to repeat an instruction. There are 4 (four) instructions, meaning 4 tasks:

(1) Please <u>unlock the phone</u>. This can be done via the screen, by the button located on the right of the device, or by the center button. The unlock method for this specific device is a numeric code. To unlock the phone the code is 2584. Once the task is completed, we will confirm its completion.

(2) <u>Write a message</u>. On the screen, you'll find three apps used for people to communicate. Identify the one that you can use to send a message to the contact "Peter". The message should say: "Peter, I'll go to the pharmacy at noon.". Once the task is completed, hit send, and we will confirm its completion.

(3) Look for a photography application. <u>Take a photo</u> centered on the orange ball that's right there (pointing) behind you. Once the task is completed, we will confirm its completion.

(4) After finishing, <u>turn off the phone</u>. The test will be concluded after switching off the device, by pressing and holding the button located on the right of the device, and then clicking on 'power off', not just locking the screen. The device must be turned off."

### INSTRUCTIONS FOR TEST ADMINISTRATORS

(1) After reading these instructions, hand over the device to the participant/respondent. (2) After this read, hand it over to him/her as quickly as possible and start counting the time. (3) Remember to announce loudly (or through high-volume automated audio) the completion of each task. Say only: "Task completed". (4) Don't forget to note down the total time and the time for each task (a), the number of instruction readings (b), and which tasks were completed (completed, yes or no) (c). (5) Automate the timing as much as you can to avoid errors and record the test, with audio and video, whenever possible.